# SDPD Round Robin 2002 Results

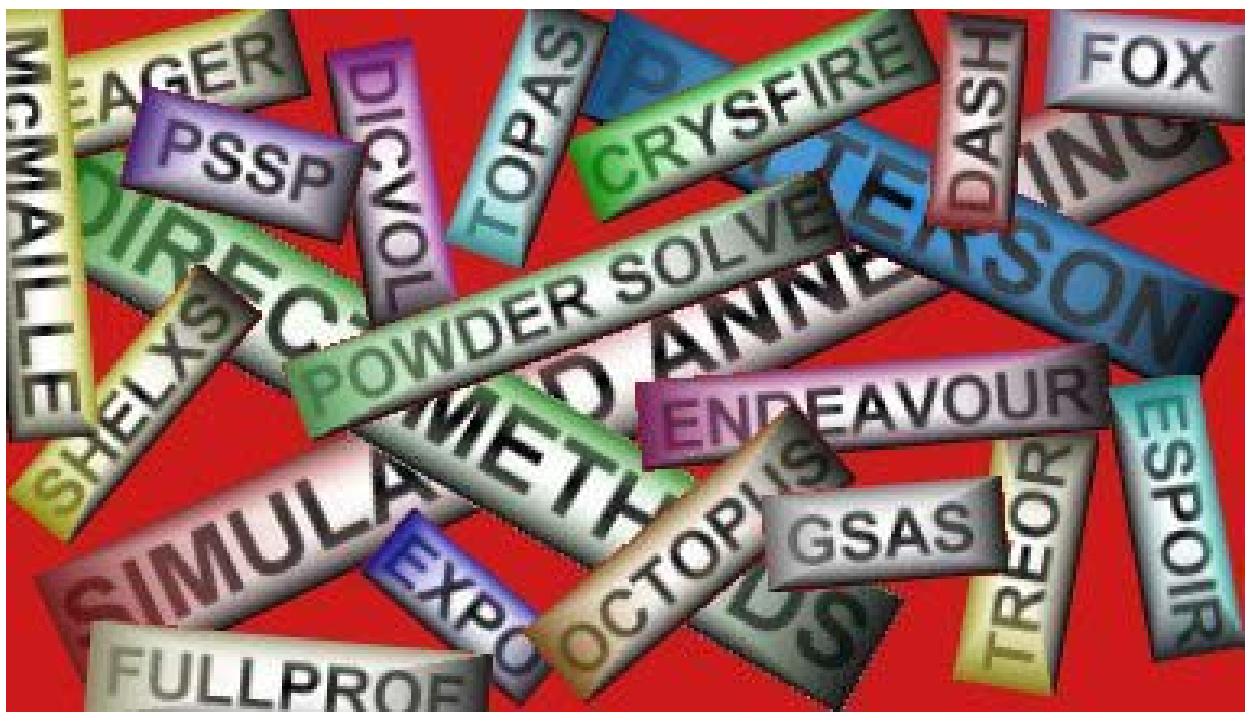

**Armel Le Bail (1) and Lachlan Cranswick (2)**
(1) *Université du Maine, CNRS UMR 6010, Avenue O. Messiaen, 72085 Le Mans, France - E-mail: alb@cristal.org*
(2) *CCP14 - School of Crystallography, Birkbeck College, Malet Street, Bloomsbury, WC1E 7HX, London, UK - E-mail: lm.d.cranswick@dl.ac.uk*

**Abstract -** Four years after the first structure determination by powder diffractometry (SDPD) round robin, a second one shows that SDPD is still not routine.

## Introduction

The total number of structure determinations by powder diffractometry (SDPD) is now more than 600. It was almost doubled during the last 4 years, since the first 1998 SDPD round robin (RR) [1-2]. Four years ago, the routine solution of 15-30 non-hydrogen atom structures from powder data was not demonstrated at the SDPDRR-1. If more than 70 potential participants downloaded the data, only two solutions were provided (by using the *DASH* program and the *CSD* package - software descriptions and Internet hyperlinks can be found at the CCP14 [4]) for the pharmaceutical sample (~30 non-H atoms), and none for the inorganic one (15 non-H atoms). Four years later, we observe that the number of structures solved by SDPD remains below 100 per annum, in spite of an explosion of the number of different software available for powder structure solution, commercial as well as in the public domain or even open source (see a random uncomplete software list in the logo above, suggesting that a user could be lost in the maze and could ignore which button to press). By the end of the XIXth IUCr congress (Geneva, August 2002), it was thus considered timely to verify if SDPD on demand had really become routine, meaning that software in other hands than those of their developers would produce solutions with some regularity and ease.

## Samples and schedule

The first SDPD Round Robin was made knowing the cell and space group parameters. Thus, the indexing "bottleneck" was avoided. A few IUCr 2002 Geneva congress attendees were expecting for an indexing round robin, The SDPDRR-2 was thus organized in two steps [3] :

**1-** Indexing - September 9 to October 13, 2002 - 8 powder patterns
**2-** Structure solution and refinement - September 9 to November 17, 2002 - 3 powder patterns

The results of the indexing step were disclosed October 14, 2002, because experts in indexing could be different from experts in structure solution. Owing to the presupposed larger difficulty at step 2, only the first three of the eight powder patterns to be indexed at step 1 had to be considered for the structure solution step 2. The sample description was the following :

**Sample 1 -** probable formula $Al_2F_{10}[C_6N_4H_{20}]$ - probable impurity : pyrochlore structure, cubic, Fd3m, a ~ 9.8 A, with possible formula $Al(F,OH)_3.xH_2O$ - Cartesian coordinates provided for the $C_6N_4H_{20}$ molecule, as found in $C_6N_4H_{20}.Cl_4$ - crystal chemistry : the usual behaviour of Al in F environment is to form $AlF_6$ octahedra - reflection laboratory X-ray data from a Bruker D8 Advance diffractometer, CuKalpha, recorded from a micro sample (40 mg).
**Sample 2 -** probable formula $Sr_5V_3(F/O/OH/H_2O)_{22}$ - capillary synchrotron data, $\lambda$ = 0.79764 Å.
**Sample 3 -** probable formula $C_{61}Br_2$ - Cartesian coordinates provided for the $C_{60}$ molecule - capillary synchrotron data, $\lambda$ = 0.79764 Å.
**Sample 4 -** probable formula $C_{20}H_{12}O_6$ - capillary synchrotron data, $\lambda$ = 0.79776 Å.
**Sample 5 -** probable formula $C_{16}H_{26}O_6$ - capillary synchrotron data, $\lambda$ = 0.79776 Å.

**Sample 6 -** probable formula $C_{21}H_{15}Bi_2O_9$ - capillary synchrotron data, $\lambda = 0.79764$ Å.
**Sample 7 -** probable formula $C_7H_5BiO_4$ - capillary synchrotron data, $\lambda = 0.79764$ Å.
**Sample 8 -** probable formula $Rb_7Cr_6F_{25}$ - reflection laboratory X-ray data from a Bruker D8 Advance diffractometer, CuKalpha.

The hunt for accurate peak positions is software- and user-dependent. The round robin participants were encouraged to use the software of their choice, though a list of peak positions was provided for each sample, obtained by using *PowderX* in default peak finding mode. At step 1, we expected only one cell proposal from every participants, for each sample (or at least for samples 1-3 from people making SDPDs routinely), together with the list of indexed peaks, and figures of merit, the name of the software (for indexing and peak position hunting), a statement telling if the cell was ascertained by applying *CHEKCELL* or/and a whole pattern fitting decomposition program (if yes, the software name and the method name - Pawley or Le Bail fit, etc), and finally a space group proposal (telling if estimated manually or with a software help, giving the software name, if any), plus explanations judged necessary (use of databases like CSD, ICSD, ICDD, etc). However, giving a list of possible unchecked cells ordered by decreasing value of figures of merit was accepted too. Robin Shirley passed on CDT files (converted from the *PowderX* peak list) ready for using the *CRYSFIRE* suite (which links to over eight indexing programs). One can note that a new Monte Carlo indexing software (*McMaille*) was born during the round robin and that improvements of *Index*, by Joerg Bergmann, were proposed. For step 2, all details of the structure solution and refinement were expected to be provided by the participants into a form available at the SDPDRR-2 Web site [3]. Slightly more than 100 participants downloaded the data.

## Results and discussion

A citation is appropriate, from a chapter of the recent book **Structure Determination from Powder Diffraction Data** [5] : "*Sadly, despite the excellent quality of synchrotron data and the ingenuity of the scientists who collect it, not all structures can be solved. Every laboratory has a supply of powder data sets which stubbornly refuse to yield to structure solution. It is the existence of these sets that will act as a continual challenge to both instrumental and algorithmic developments*." We completely subscribe to that statement. It suggests that the more than 600 structures determined *ab initio* from powder diffraction during the last 20 years are the tip of the iceberg. SDPD is known to be quite difficult, but scientists are secret about their failures, so that SDPD could even be much more difficult than usually said. It was thus decided to include in this second SDPD round robin some of those (splendid) "*data sets which stubbornly refuse to yield to structure solution*." These are samples 4-8 for which we had no structure solution, so that the cell proposals themselve cannot be confirmed. Of course, some samples having known (unpublished) solutions were also incorporated (samples 1 to 3).

**Step 1 : The indexing bottleneck -** Six participants sent results in due time for the step 1, giving generally cell parameters for sample 1 to 3. Only one participant suggested cell parameters for some samples (5, 7 and 8) of the supplementary series 4 to 8. These contributions appeared so scarce (6% return rate) that it was announced that post-deadline proposals will be accepted.

**Table 1 - Participant (Pn) cell proposals to samples 1-8 (Sn).**

| | **S1** | **S2** | **S3** | **S4** | **S5** | **S6** | **S7** | **S8** |
|---|---|---|---|---|---|---|---|---|
| **P1** | X | X | X | | | | | |
| **P2** | X | X | X | | | | | |
| **P3** | X | X | X | | | | | |
| **P4** | | X | | | | | | |
| **P5** | X | X | X | | X | | X | X |
| **P6** | X | X | X | | | | | |

For samples 1-2, all the returned cell proposals were correct (sample 1, monoclinic : $a$ = 10.323 Å, $b$ = 7.395 Å, $c$ = 8.535 Å, $\beta$ = 91.29°; sample 2, monoclinic : $a$ = 11.239 Å, $b$ = 8.194 Å, $c$ = 19.943 Å, $\beta$ = 106.727°). Sample 3 is cubic ($a$ = 18.88 Å), it was proposed by three of the five participants having sent a cell proposal, while the two remaining ones proposed a tetragonal subcell (possibly due to default volume restrictions in the software used - but an application of the LePage algorithm via *Chekcell* should have revealed the error). Software or program suites used were *CRYSFIRE* (P1, P3), *DICVOL* (P2), *ITO* (P4), *Index* (P5) and *X-Cell* (P6).

For samples 4 to 8, only one response was obtained in due time for three of the five data - using the *Index* software of Joerg Bergmann [6]. More Post-deadline results were sent by participants 5 and 6 (the latter using the *X-Cell* software [7]). In the absence of any crystal structure, we cannot confirm nor invalidate these cell proposals. These samples 4-8 are part of the great mass of iceberg under the surface.

**Table 2 - Post-deadline cell proposals to samples 1-8 :**

| | S1 | S2 | S3 | S4 | S5 | S6 | S7 | S8 |
|---|---|---|---|---|---|---|---|---|
| **P5** | X | X | X | X | X | | X | X |
| **P6** | X | X | X | X | X | X | X | X |

It appears that powder indexing is not easy and not routine. It is known since three decades that bad data are the main cause of indexing failures. "Bad" data means both inaccurate and impure. Efforts are made in some indexing software in order to be more efficient with at least a two-phases mixture, since inaccuracy can be fought by more efforts at the recording step for well crystallized samples. However, the main problem is that an indexing software user may not have the information that the data are either inaccurate or impure or both.

Nowadays, indexing is not only using an indexing software which may produce several different cells with neighbouring figures of merit. The whole suite of operations after recording a powder pattern and after failing to identify any known compound (search/match process [8]) is (under brackets are given the techniques/software selected by the participants) : peak search [*PowderX* - as provided with the round robin data; *WinPlotr*; *CMPR*], indexing *sensu stricto* [see list above], cell reduction [*LePage*], visual checking and more [*Chekcell*], whole pattern fit checking [*Fullprof*; *Rietica*; *Materials Studio*; Pawley/Le Bail methods], space group proposal [*Chekcell*, *Materials Studio*, manual]. The cell proposals have to be checked against the whole pattern, not only the 20-30 first peak positions, and this produces cell parameters refinement as well as more data for estimating systematic extinctions from which one or several possible space groups are deduced. The next step (structure solution) can be quite time-consuming, so that a powder diffractionist will not make any further effort if the probability to have found the true cell and space group is not high (this is a question of self-convincing, as indexing is still considered as an art).

**Step 2 - Structure solution and refinement** - Two participants sent successful results for samples 1 and 2 (none for sample 3). This is no more than for the SDPDRR-1 in 1998. The software used are *FOX* (P2) [9] and *TOPAS* (P7) [10].

**Table 3 - Participant (Pn) structure proposals to samples 1-3.**

| | S1 | S2 | S3 |
|---|---|---|---|
| **P2** | X | X | |
| **P7** | X | X | |

Participant 2 solved the sample 1 structure by using *FOX* according to the following strategy : structural units location in direct space by simulated annealing in the parallel tempering mode. Three independent entities (two $AlF_6$ octahedra and the $C_6N_4H_{20}$ molecule) were allowed to rotate and move in the P2/c space group, corresponding to 18 degrees of freedom. The final refinement was made by using the *FULLPROF* software. Much more details are given by the participants at the SDPDRR-2 Web site [3].

Participant 7 solved the sample 1 structure by using *TOPAS* according to the following strategy : molecule location in direct space, simulated annealing, structure determination using step intensity data, starting with 3 "rigid" bodies (including 2 for $AlF_6$ octahedra). For the $C_6N_4H_{20}$ molecule, 3 rotational and 3 translational degrees of freedom and 4 torsion angles were used.

Participant 2 solved the sample 2 structure with *FOX* in a similar way as for sample 1 : either 3 tetrahedra $VO_4$ and 5 cubes $SrO_8$ or 3 octahedra $VO_6$ and 5 cubes $SrO_8$ were rotated and translated (48 degrees of freedom in both cases). Both models have yielded the same correct positions for the cations. The positions of anions were determined by third modeling using free atoms.

Participant 7 solve the sample 2 structure with *TOPAS* by putting into the cell 30 independent atoms (no constraint/restraint) at random positions, locating them by simulated annealing.

Computer programs used by the organizers for the step 1 (indexing) were *TREOR*, *DICVOL*, *ITO*, *CRYSFIRE*, *McMaille*, and for step 2 (structure solution) *SHELXS* (sample 2) and *ESPOIR* (samples 1 and 3). Sample 1 was solved in the Pc space group by using 3 objects in the *ESPOIR* program (one $AlF_6$ octahedron, one $AlF_4$ square plane and the $C_6N_4$ molecule - fig. 1). *SHELXS* solved the sample 2 structure by direct methods, quite easily. *ESPOIR* could also solve the sample 2 structure in "scratch" mode (all atom random at the beginning, going to their right place by Monte Carlo - fig. 2). Sample 3 could not be "solved" unless a global scattering factor for a sphere was used for modelling a $C_{60}$ disorder. The Br atoms appeared located partially at 2 positions in the I-43m space group (fig. 3).

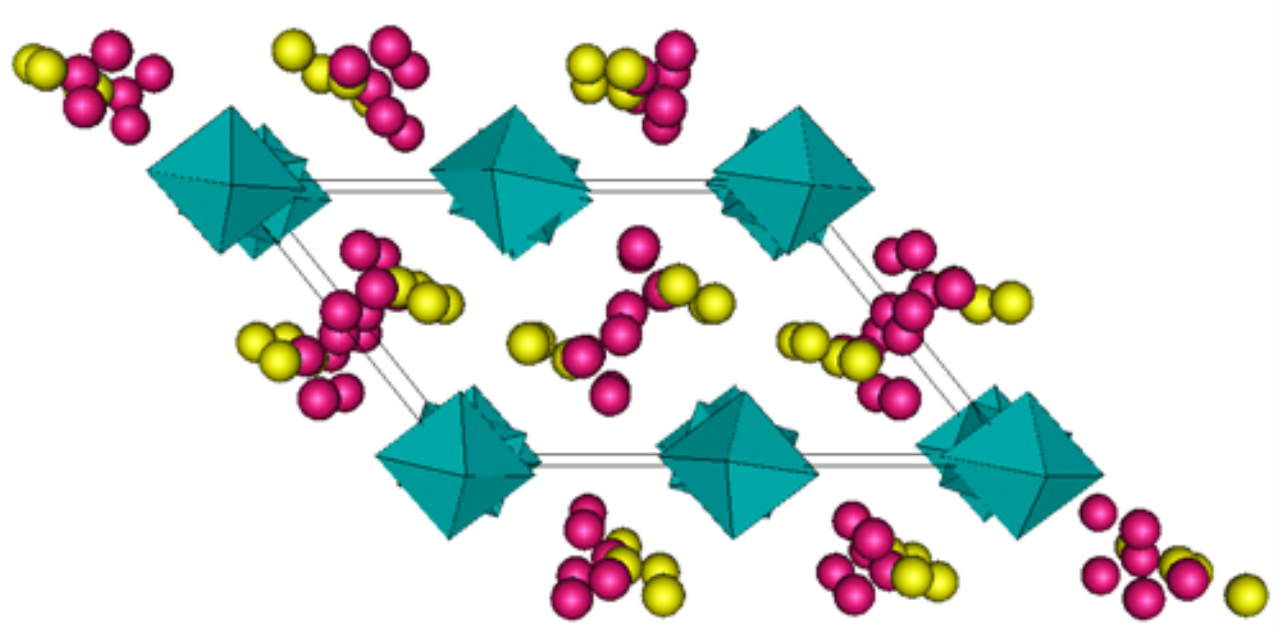

Fig. 1 - Projection of the sample 1 structure along the b axis (H atoms not represented)

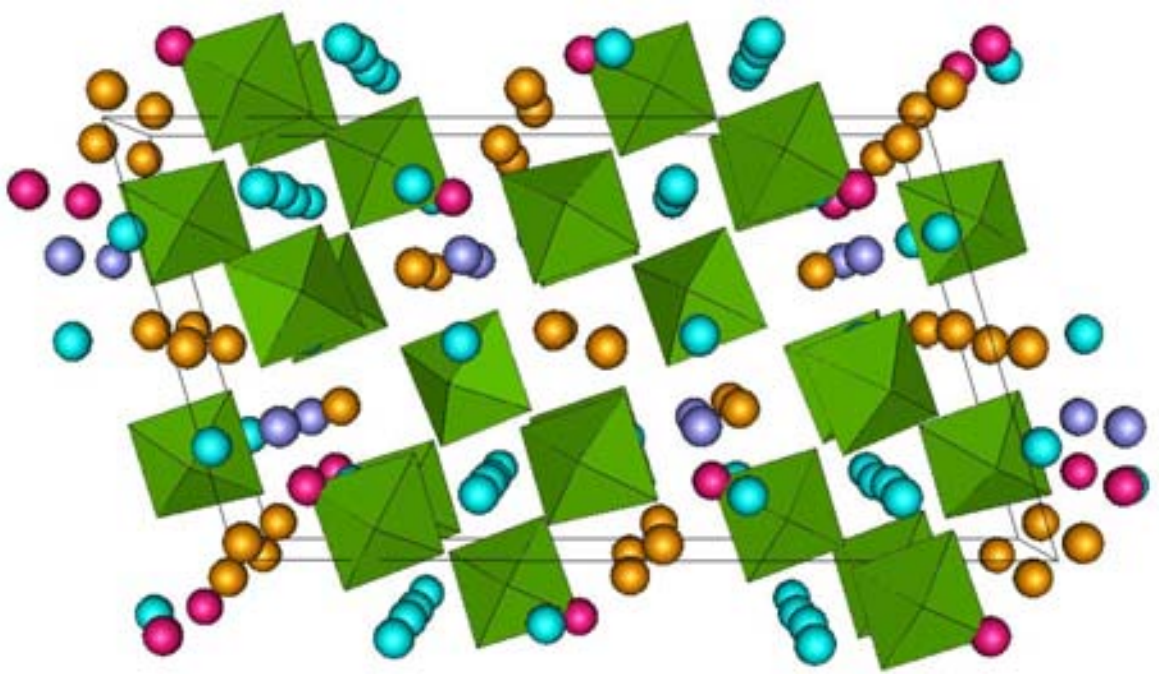

Fig. 2 - Projection of the sample 2 structure along the b axis.

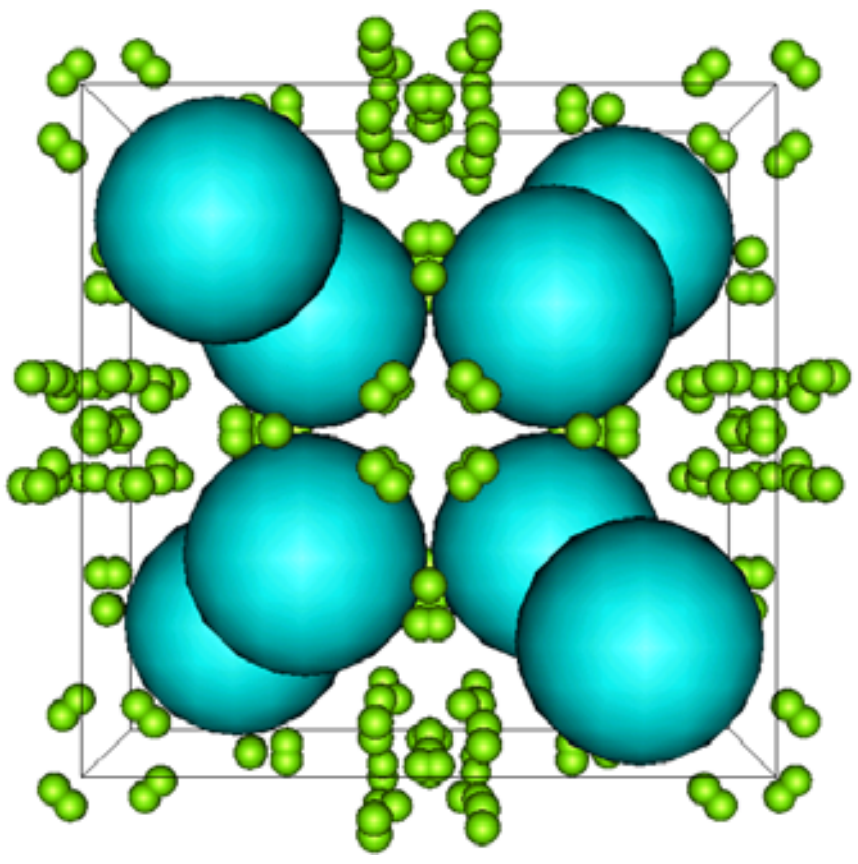

Fig. 3 - The $C_{60}CBr_2$ mean structure model for sample 3. Big spheres represent the $C_{60}$ molecules. Small spheres are the Br atoms.

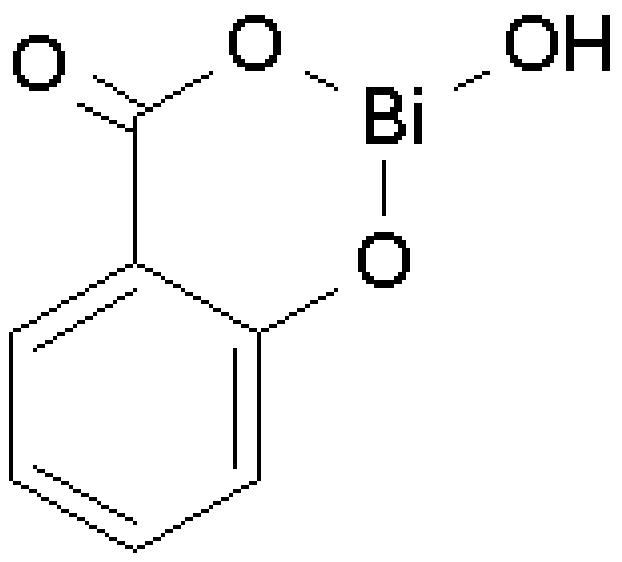

Fig. 4 - Molecular formula for sample 7.

The SDPDRR-2 is still open for post-deadline structure solution results for samples 4-8. More details can be found on these samples which could help. For instance, sample 7 is a commercial bismuth sub-salicylate [CAS : 14882-18-9] whose formula $C_7H_5BiO_4$ can also be found written as : 2-(HO)$C_6H_4CO_2$BiO (pepto-bismol, a drug for the traveler's diarrhea - can give you a black tongue). It was shown (together with sample 3) at IUCR XVIII, Glasgow, 1999 in a poster comparing the merits and advantages of synchrotron versus conventional X-ray [11]. Its molecular formula appears quite simple (fig. 4).

## Conclusion

The two months allocated for solving at least one of 3 structures by SDPD methodology would have been quite enough if SDPD was routine (a rainy evening would have been sufficient). Only two final answers for samples 1 and 2 suggest that either SDPD experts are lazy or SDPD is still not routine. This will be the main SDPDRR-2 conclusion. Anyway, three computer programs certainly deserve special mention because they could provide exceptional results in due time : *Index* for indexing (participant 5) and *FOX* (participant 2) as well as *TOPAS* (participant 7) for structure solution. The SDPDRR-2 seems to establish the triumph of direct space methods (*FOX*, *TOPAS*). But statistics cannot seriously be deduced from two contributions (owing to the number of data downloads, the return is less than 2%). Considering that still less than 100 SDPDs are published per year, it was clearly utopian to expect that 3% of them could be solved in 2 months by people extremely busy at solving their own difficult problems. The final examination of the SDPD Internet Course [12] is a complete SDPD, the subscribers having two weeks for the structure determination of a moderatly complex case. Some are able to finish the job in less time (4-5 days). Since subscribers at the end of the course become experts, this may define the mean time needed for a simple SDPD. That prohibiting time of several days may lead to a choice : either participants solve their own problems or they solve the round robin problems. Ordinary users could have preferred to solve their own problems. But software developers have other motivations which are to promote their product, in principle, so that wasting days to solve the round robin problems may appear worthwhile to them. Is the absence of software like *EXPO*, *DASH*, *POWDERSOLVE*, *PSSP*, *MRIA*, *FOCUS*, *SIMULATED ANNEALING*, *ENDEAVOUR*, *EAGER*, *OCTOPUS*, *SAFE* (etc) indicative of failure to solve or of lack of willing to solve ? We do not have answers.
For those wanting to continue to play with the SDPDRR-2 data sets, they will remain available [3], in the hope that they will not definitely stay in a category "*which stubbornly refuse to yield to structure solution*". Anyway, a doubling of the total number of SDPDs is expected again during the next four years, this will have to correspond to a new revolution that would allow 200 structures to be determined per year. A rendez-vous is given for the SDPDRR-3 in 2006, expecting for more successful participants than just two or three out of over a hundred participants.